# Odd modes of Kelvin–Helmholtz instability in sheared flows of plasmas and fluids

A.Yu. Chirkov,[1] V.I. Khvesyuk

Bauman Moscow State Technical University, Moscow 105005, Russia
[1]e-mail: alexxeich@mail.ru

The common feature of sheared flows of an ideal fluid and plasma in magnetic field is the Kelvin–Helmholtz instability. This instability is described by identical equations in mentioned two cases. The wave equation for the eigenmodes in the plasma obtained by the kinetic method in the long-wave limit coincides with the Rayleigh equation known for an ideal fluid. Velocity profile with a transition layer of finite width is considered. Odd modes are investigated. Localized velocity shear layer is the source of perturbations propagating beyond the transition layer. The conditions of localization of the modes are studied.

## 1. Introduction

Kelvin–Helmholtz instability is a common feature of sheared flows of fluids and magnetized plasmas. In an ideal fluid, this instability is described by Rayleigh equation [1]. The problem of Kelvin–Helmholtz instability in the plasma confined by the magnetic field is important for theories of the so-called transport barriers [2, 3]. Transport barriers appear in the localized regions of the plasma column due to strong shear of the plasma flows across magnetic field lines. Growing velocity gradient decreases the level of the drift-wave driven plasma turbulence in the transport barrier, but on the other hand it excites the Kelvin–Helmholtz instability that limits the effectiveness of the drift turbulence suppression.

Plasma microturbulence characterized by extremely small perturbations of the hydrodynamic fields which distinguishes it from classical turbulence in fluids [4]. Plasma oscillations are transmitted not by direct collisions of particles, and they are transferred by electromagnetic fields. The linear stability approach to plasma wave modes [5] based on the method of small perturbations as well as it in the case of the fluid [1, 6, 7]. But in the case of plasma the essential difference is the use of kinetic analysis since the hydrodynamic approximation does not allow concluding about stability in many cases. Transport processes in the plasma caused by microturbulence which formally can be analyzed on the basis of the classical kinetic description of turbulence [8]. Stability theory of fluid flows preceded the theory of plasma instabilities and largely formed the basis of the latter. Theory of microturbulence currently developed for stationary sheared flows of fluids. In the case of the plasma, kinetic methods of analysis of instabilities reached the highest development. These methods take into account the resonant excitation mechanisms based on the motion of individual particles.

Kinetic theories of transport as well as spectral theories were developed to calculate macroscopic quantities from the spectra of turbulence. Turbulent transport is often corresponds to superdiffusion regime which is observed, for example, in a plasma in a magnetic field [9, 10]. A phenomenological approach can be used for nondiffusive transport of particles, whereby the flow of particles is determined on the basis of Maxwell–Cattaneo–Vernotte equations [11, 12]. On the other hand, the problem of determining the turbulent transport fluxes is solved by using models based on direct numerical simulations [13] which require no additional equations for the relaxation.

Models of turbulence excitation in fluid flows developed now [14, 15]. Models of two-dimensional turbulence in a fluid developed also [15, 16]. From the standpoint of the general laws of fluids and plasmas it is important as two-dimensional turbulence is typical phenomena for the plasma in magnetic field. According to the kinetic approach, the mechanisms of turbulence generation are largely dependent on the conditions for the excitation of the instabilities. On the stage of growth of perturbations their dynamics can be considered on the basis of linear analysis using linear growth rate. On the strongly nonlinear stage the amplitudes of the initial small perturbations reach a level at which the dissipative mechanisms cause the decay of perturbations.

Section 2 discusses the formulation of the linear analysis problem of Kelvin–Helmholtz instability which is required to determine the marginal stability conditions and the influence of the boundary conditions on stability. We consider two-dimensional perturbations as they have a lower threshold of instability than the three-dimensional disturbances according to the Squire theorem. In this paper we are interested in the odd modes. In the Section 3, calculations results are presented. Section 4 contains the conclusions.

## 2. The model

For an ideal incompressible fluid the initial system of equations includes the continuity equation and Euler equations. We suppose that the flow moves along the $y$-axis with the unperturbed velocity $V(x)$ depending only on the coordinate $x$. Perturbed components of the velocity are given by the stream function $\psi(x,y)$: $v_x = \partial \psi / \partial y$, $v_y = -\partial \psi / \partial x$. As a result of linearization of the original system of equations it is reduced to a single equation describing the perturbed stream function. In this case, it can be written as

$$\psi(x, y) = \varphi(x)\exp(-i\omega t + iky), \qquad (1)$$

where $\omega$ is the complex frequency, $k$ is the wave number, function $\varphi(x)$ describes transversal profile of the perturbation (amplitude variation across the unperturbed flow).

The function $\varphi(x)$ satisfies the Rayleigh equation

$$\frac{d^2\varphi}{dx^2} + \left(-k^2 + \frac{d^2V/dx^2}{\omega/k - V}\right)\varphi = 0. \qquad (2)$$

This equation and appropriate boundary conditions are eigenvalue problem, where $\varphi(x)$ means eigenfunction for eigenvalue $\omega$, and $k$ is the wave number of the corresponding mode.

In plasma, Kelvin–Helmholtz instability appears in the presence of nonuniform $\mathbf{E}\times\mathbf{B}$-drift in crossed electric ($\mathbf{E}$) and magnetic ($\mathbf{B}$) fields [17–21]. In particular, this instability was observed experimentally in cylindrically symmetric configurations [20].

The wave equation of the electrostatic Kelvin–Helmholtz instability in the plasma has been obtained in the framework of the kinetic approach [19]. In terms of large-scale violations of the stability of the plasma configuration, the long-wavelength hydrodynamic limit is the most important. So we consider hydrodynamic limit of the kinetic formulation of the problem [22]. The condition for establishing the this limit is the following: $k\rho_{Ti} \ll 1$, where $\rho_{Ti}$ is the thermal cyclotron radius of ions in a magnetic field. In the case of two-dimensional hydrodynamic limit, kinetic wave equation [19] coincides with the Rayleigh equation (2). Functions $\psi(x, y)$ and $\varphi(x)$ correspond to the electrostatic wave potential in the plasma. In the hydrodynamic limit, the perturbed velocity is associated with the **E**×**B**-drift in the perturbed electric field of the wave. Magnetic field is supposed to be uniform. Under these assumptions, functions $\psi(x, y)$ and $\varphi(x)$ also describe the velocity streamlines. Function $V(x)$ is the unperturbed flow velocity of the plasma (**E**×**B**-drift) which is similar the flow velocity of the fluid. Thus the long-wave perturbations in the plasma are completely analogous with the ideal fluid case. According this analogy, the results obtained on the basis of the Rayleigh equation (2) can be applied to an ideal fluid as fare as for the plasma in a magnetic field.

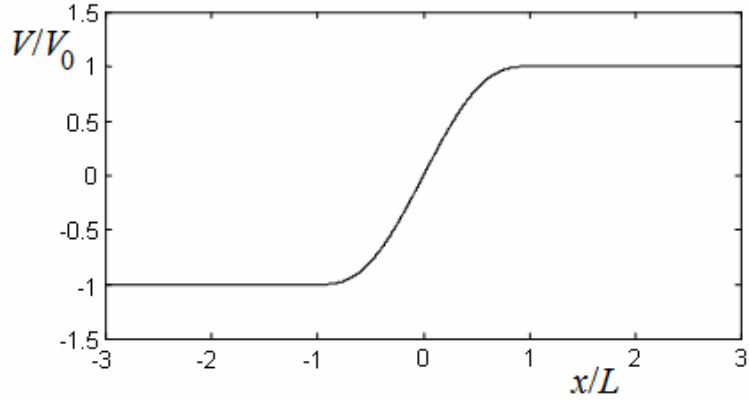

Fig. 1. The unperturbed velocity distribution in the counter-flow configuration.
The with of transition layer is $2L$

Consider a continuous profile of the unperturbed velocity $V(x)$ with a change in a limited layer of width $2L$ (see Fig. 1). The first and second derivatives are also continuous at any $x$. The direction of the $x$-coordinate is chosen so as shown in Fig. 1. This approximation corresponds to the two opposing flows with the same velocity $V_0$ outside the transition layer. The analytical expression for the velocity distribution is as follows:

$$\frac{V(x)}{V_0} = \begin{cases} -1, & \dfrac{x}{L} < -1 \\ \dfrac{15}{8}\dfrac{x}{L} - \dfrac{5}{4}\left(\dfrac{x}{L}\right)^3 + \dfrac{3}{8}\left(\dfrac{x}{L}\right)^5, & -1 \leq \dfrac{x}{L} \leq 1 \\ 1, & \dfrac{x}{L} > 1 \end{cases} \qquad (3)$$

The perturbed velocity components are associated with the function $\varphi(x)$ by the relations

$$v_x = \partial \psi / \partial y = ik\varphi(x)\exp(-i\omega t + iky), \qquad (4)$$

$$v_y = -\partial \psi / \partial x = -(d\varphi/dx)\exp(-i\omega t + iky). \qquad (5)$$

Here we consider only the odd modes, i.e. $\varphi(x)$ is odd function. We use the following conditions for the odd function $\varphi(x)$: $\varphi = 0$ and $d\varphi/dx = 1$ at $x = 0$. Solutions are analyzed for several variants of the boundary conditions listed below.

The boundary conditions of the first kind: $\varphi = 0$ ($v_x = 0$) at $x = \pm H$, where $H$ is the half-width of the localization of the mode (generally $H \neq L$). In particular, for the fluid flow this condition can be satisfied, if solid walls exist. And the demand $d\varphi/dx = 0$ ($v_y = 0$) on the walls is not required.

The boundary conditions of the second kind: $d\varphi/dx = 0$ ($v_y = 0$) at $x = \pm H$. Such boundary conditions can not be posed on the walls as the transverse velocity at the wall can be nonzero. Fluctuations in this case should be allowed to exist in the region $|x| > H$.

The boundary conditions of mixed type: $\varphi = 0$ ($v_x = 0$) and $d\varphi/dx = 0$ ($v_y = 0$) at $x = \pm H$. Such boundary conditions are redundant. In general, they are unlikely to be performed. However, these conditions are acceptable, if $H \to \infty$. Therefore the corresponding modes can take place in an unlimited range of $x$.

To present the results of calculations in dimensionless form we choose the following scales: velocity $V_0$, length $L$, frequency and growth rate $\omega_0 = V_0/L$. Taking into account the symmetry of the problem it is easy to see that in the chosen coordinate system, the wave phase velocity is equal to zero and therefore $\text{Re}(\omega) = 0$. Thus the eigenvalues of the problem are purely imaginary $\omega = i\gamma$, where $\gamma = \text{Im}(\omega)$ is the growth rate of the mode.

## 3. Results of the calculations

Eigenvalues, eigenfunctions, and the stability regions were obtained as a result of numerical calculations. Algorithm for numerical solutions was developed on the basis of the Runge–Kutta methods. In Fig. 2, the dimensionless growth rate $\gamma/\omega_0$ versus the dimensionless wave number $kL$ is presented for the instability of unbounded modes. As one can see, if $kL < 0.1$, this dependence is close to linear law $\gamma/\omega_0 \approx 0.5kL$. Growth rate reaches maximum value $\gamma \sim 0.1\omega_0$ at $kL \approx 0.25$. The instability region is bounded above by a wave number $kL \approx 0.4$.

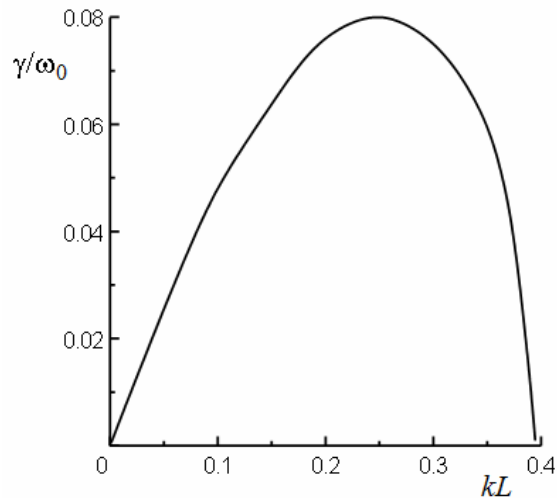

Fig. 2. Dimensionless growth rate $\gamma/\omega_0$ versus dimensionless wave number $kL$ for modes in an unbounded flow

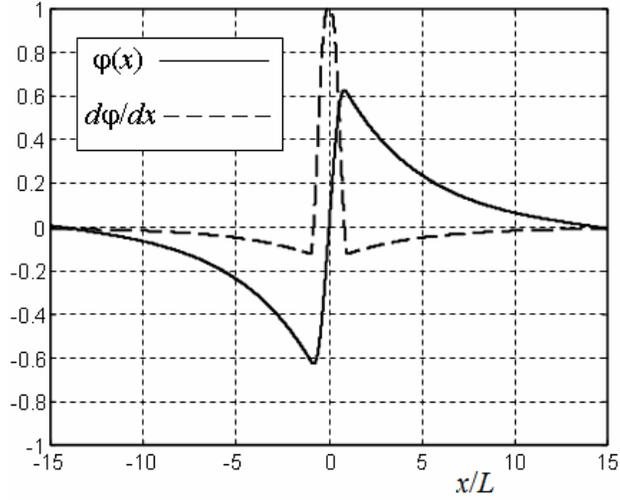

Fig. 3. Eigenfunction $\varphi(x)$ (solid line) and its derivative $d\varphi/dx$ (dashed line) for mode with $kL = 0.2$ (growth rate $\gamma = 0.076\omega_0$)

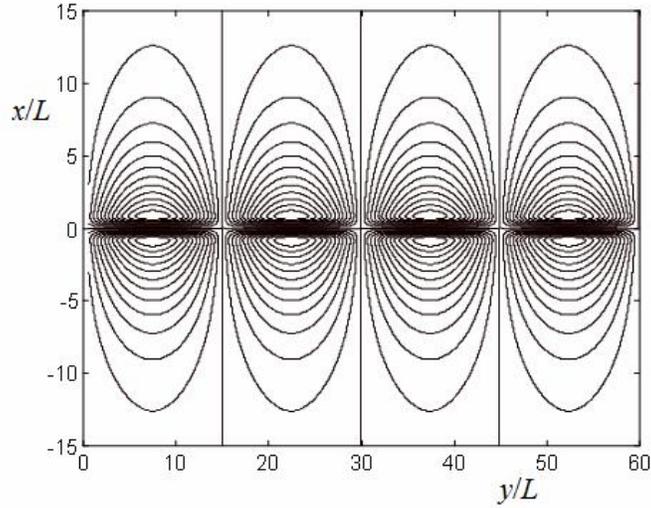

Fig. 4. Contours of perturbed stream function $\psi(x, y)$ of the mode with $kL = 0.2$ and $\gamma/\omega_0 = 0.076$

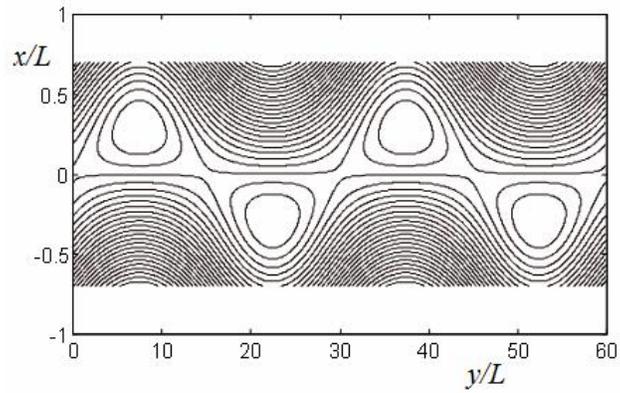

Fig. 5. Vortices near the inflection point ($x = 0$) formed by the stream function $\psi_\Sigma(x, y)$. The mode parameters are as follows: $kL = 0.2$ and $\gamma/\omega_0 = 0.076$. The maximum amplitude of the perturbed velocity is assumed to be $0.5V_0$

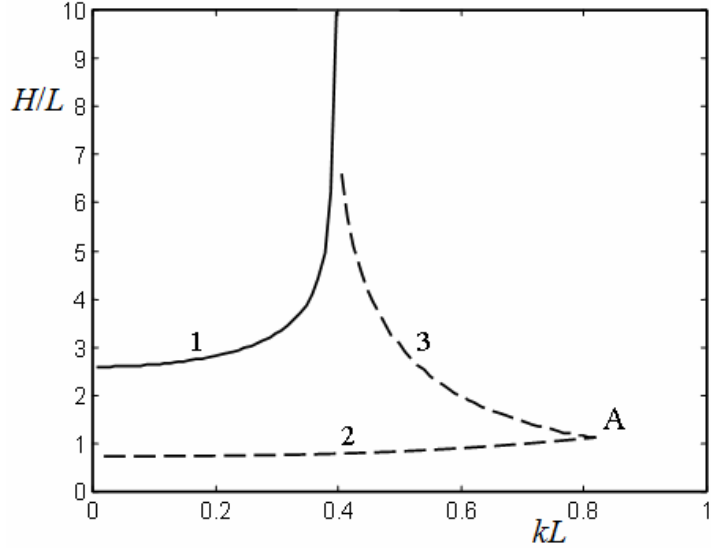

Fig. 6. The boundaries of the stability regions corresponding to the boundary conditions of the first kind (solid line 1) and second kind (dashed lines 2, 3)

Eigenfunctions $\varphi(x)$ are shown in Fig. 3 which also shows the derivative $d\varphi/dx$. The contours of the corresponding perturbed stream function $\psi(x, y)$ are shown in Fig. 4.

Fig. 5 shows the vortex flow structure in the region near the inflection point of $V(x)$ emerging with the development of a single mode of instability. In Fig. 5, the stream function $\psi_\Sigma(x, y)$ includes the stream function of the unperturbed flow (velocity profile given by (3)), and the perturbed stream function $\psi(x, y)$.

As shown by calculations, in the cases of boundary conditions of the first and second kinds, the typical values of the maximum growth rates are the same order of magnitude as for the modes in an unbounded flow. A qualitative picture of solutions within the transition layer is also analogous to the unrestricted flow. For the considered modes one can set the boundary of stability and find the mode width $H$ corresponding $\gamma \to 0$. Note that our numerical algorithm can not be used for strictly $\gamma = 0$. So the boundary curves are results of the growth rate numerically approaching zero with a given accuracy.

The results of calculations of the boundary curves are shown in Fig. 6. The region of instability ($\gamma > 0$) for the modes obtained with the boundary conditions of the first kind located above the curve 1 in Fig. 6. When considering the boundary conditions of the second kind, two zeros of the derivative $d\varphi/dx$ can be found (curves 2 and 3). The position of the first zero of the derivative (curve 2) slightly varies with the wave number, as it is determined by the structure of the flow inside the transition layer. For modes corresponding to second zero of the derivative $d\varphi/dx$ the region of instability is above the curve 3 to the left of point A.

## 4. Conclusions

This work presents an analysis of Kelvin–Helmholtz instability in sheared flows of fluids and magnetized plasmas. In the long-wave limit the equation for the eigenmodes for the plasma obtained by the kinetic method coincides with the Rayleigh equation for an ideal fluid. Stability of the velocity distribution with a transition layer of type of counter-flow was examined in our analysis. In such a flow, Kelvin–Helmholtz instability develops. The typical growth rate value is $\gamma \approx 0.1 V_0 / L$. Growth rate is approximately the same in the case of the instability in an unlimited flow of an ideal fluid with a velocity profile approximated by hyperbolic tangent [23] as well as the instability in channel with walls [24]. The growth rate of the same order is characterizing the instability in the case of a compressible viscous fluid [25]. A qualitative picture of the instability is also close to the case of a compressible plasma in a magnetic field in the approximation of ideal magnetohydrodynamics (MHD) [26].

In Ref. 27, two-fluid MHD approach with no dissipation show that density variations associated with the Kelvin–Helmholtz instability can lead to the development of secondary instabilities of hydrodynamic type. Therefore the generation of sheared flows with very large values of shear parameter $\gamma_s = dV/dx$ is undesirable from the standpoint of the Kelvin–Helmholtz instability and secondary instabilities. These instabilities lead to mixing which dramatically increases the intensity of transport. On the other hand, the generation of shear flows is the main way to reduce the turbulent transport in the plasma in a magnetic field.

In our opinion, the results of this study are necessary to further analysis of microturbulence based on nonlinear dynamics of perturbations of finite amplitude. Such methods opens the possibility to develop self-consistent theory of turbulent transport in different media.

## Acknowledgments

This work was supported by RFBR, project No. 11-08-00700-a.

## References

[1] C.C. Lin, The theory of hydrodynamic stability, Cambridge University press, 1955.
[2] R.C. Wolf, Plasma Phys. Contol. Fusion **45** (2003) R1–R91.
[3] J.W. Connor, T. Fukuda, X. Garbet, et al., Nucl. Fusion **44** (2004) R1–R49.
[4] P.G. Frick, Turbulence: approaches and models, Regular and Chaotic Dynamics, Moscow–Izhevsk, 2003.
[5] A.B. Mikhailovskii, Theory of plasma instabilities, Consultants Bureau, New York, 1974.
[6] P. Drazin, Introduction to hydrodynamic stability. Cambridge University Press, 2002.
[7] P.G. Drazin, W.H. Reid, Hydrodynamic stability. Cambridge University Press, 2004.
[8] C.C. Lin, in Turbulent Flows and Heat Transport, Vol. V, Chapter III, Princeton University Press, 1959.
[9] B.P. Van Milligen, E. de la Luna, F.L. Tabares, et al., Nucl. Fusion **42** (2002) 787–795.
[10] D. Del-Castillo-Negrete, B.A. Carreras, V.E. Lynch, Phys. Plasmas **11** (2004) 3854–3864.
[11] D. Jou, J. Casas-Vázquez, G. Lebon, Extended irreversible thermodynamics, Springer, 2001.
[12] F. Vázquez, F. Márkus, Phys. Plasmas **17** (2010) 042111.
[13] V.I. Khvesyuk, A.Yu. Chirkov, Tech. Phys. **49** (2004) 396–404.
[14] G. Gioia, F.A. Bombardelli, Phys. Rev. Lett. **88** (2002) 014501.
[15] G. Gioia, P. Chakaborty, Phys. Rev. Lett. **96** (2006) 044502.
[16] N. Guttenberg, N. Goldenfeld, Phys. Rev. E **79** (2009) 065306.
[17] F.W. Perkins, D.L. Jassby, Phys. Fluids **14** (1971) 102–115.
[18] D.L. Jassby, Phys. Fluids **15** (1972) 1590–1604.
[19] G. Ganguli, Y.C. Lee, P.J. Palmadesso, Phys. Fluids **31** (1988) 823–838.
[20] J.C. Glowienka, W.C. Jennings, R.L. Hickok, Phys. Fluids **31** (1988) 2704–2709.

- 8 -
[21] A.V. Timofeev, Plasma Phys. Control. Fusion **43** (2001) L31–L35.
[22] A.Yu. Chirkov, in VII Int. symposium on radiation plasma dynamics, NIC "Engineer", Moscow, 2006, p. 186–189.
[23] A. Michalke, J. Fluid. Mech. **19** (1964) 543–556.
[24] M. Zhuang, P.E. Dimotakis, T. Kubota, Phys. Fluids **A2** (1990) 599–604.
[25] N.D. Sandham, W.C. Reynolds, J. Fluid. Mech. **224** (1991) 133–158.
[26] A. Miura, Phys. Plasmas **4** (1997) 2871–2885.
[27] A. Tenerani, M. Faganello, F. Califano, F. Pegoraro, Plasma Phys. Control. Fusion **53** (2011) 015003.